\documentstyle[prl,aps,preprint,floats]{revtex}
\tightenlines
\def\re{ \\ }
\def\avrg#1{{\langle #1 \rangle}}
\def\half{{\textstyle{1\over2}}}
\def\spose#1{\hbox to 0pt{#1\hss}}
\def\lta{\mathrel{\spose{\lower 3pt\hbox{$\mathchar"218$}}
     \raise 2.0pt\hbox{$\mathchar"13C$}}}
\def\gta{\mathrel{\spose{\lower 3pt\hbox{$\mathchar"218$}}
     \raise 2.0pt\hbox{$\mathchar"13E$}}}
\def\bcdes{Bond {\it et al.} 1994}
\def\crit{Crittenden {\it et al.} 1993a}
\def\polar{Crittenden {\it et al.} 1993b}
\def\critab{Crittenden {\it et al.} 1993a,b}
\def\bbn{Walker  {\it et al.} 1991}
\def\kolb{Kolb and Turner 1990}
\def\eturner{Maoz and Rix 1993}
\def\G4{Bardeen, Steinhardt and Turner 1983, Guth and Pi 1982,
  Starobinskii 1982, Hawking 1982}
\def\AWSta{Rubakov {\it et al.} 1982, Starobinskii 1985, Abbott and Wise 1984}
\def\Dav{Davis {\it et al.} 1992}
\def\Oth{Lucchin  {\it et al.} 1992, Salopek 1992, Liddle and Lyth 1992, Sahni
and Souradeep 1992, Lidsey and Coles 1992, Adams  {\it et al.} 1993}
\def\BE{Bond and Efstathiou 1984}

\def\dmr{Smoot {\it et al.} 1992}
\def\BE{Bond and Efstathiou 1984, Bond 1988, Efstathiou 1990}

\def\msam{Cheng {\it et al.} 1993}
\def\mit{Ganga {\it et al.} 1993}
\def\ten{Watson 1993}

\def\sask{Wollack {\it et al.} 1993}
\def\gaier{Gaier {\it et al.} 1992}

\def\ovro{Readhead {\it et al.} 1989}
\def\max{Meinhold {\it et al.} 1993, Gunderson {\it et al.} 1993}
\def\pwd{Tucker {\it et al.} 1993}
\def\python{Dragovan {\it et al.} 1993}
\def\silk{White {\it et al.} 1993}
\def\kofe{Kofman and Starobinskii 1986}
\def\kofn{Kofman {\it et al.} 1992}
\def\hope{Polnarev 1985, Ng and Ng 1993, see also M.V. Sazhin in these
Proceedings}

\begin{document}
\draft
\title{How Well Can Cosmological Parameters Be Estimated from CMB
Observations? }

\author{
J. Richard Bond,$^1$
Richard L. Davis,$^2$
Paul J. Steinhardt$^2$}
\address{
$^{(1)}$  Canadian Institute for Theoretical Astrophysics,
University of Toronto,
Toronto, Ontario, Canada
M5S 1A7  \\
$^{(2)}$  Department of Physics, University of Pennsylvania,
Philadelphia, PA  19104}
\maketitle
\begin{abstract}
The CMB anisotropy depends sensitively upon the slope and amplitude of
primordial density and gravitational wave fluctuations, the
baryon density, the Hubble constant, the cosmological constant, the
ionization history, {\it etc.}  We report on recent work showing
how well multi-scale measurements of the
anisotropy power spectrum can  resolve these factors. We identify a
hypersurface in cosmic parameter space that can be accurately
localized by observations, but along which
the likelihood will hardly vary. Other cosmic observations
will be needed to break this degeneracy.
\end{abstract}

\vspace{.2in}
\noindent
For questions, contact author Paul J. Steinhardt, University of
Pennsylvania [steinh@steinhardt.hep.upenn.edu] (215-898-5949)

\noindent
6 FIGURES, no tables

\newpage

In this paper, we discuss the degree to which the CMB anisotropy
observations can determine cosmological parameters such as the slope
of the initial power spectrum, the age of the universe and the
cosmological constant.  These proceedings are a summary of and
expansion upon a recent series of studies [\bcdes, \critab].
Our central conclusion is that CMB anisotropy measurements alone
cannot fix the parameters individually; however, a non-trivial
combination of them can be determined.  More concretely, for models
based on the generation of gaussian, adiabatic fluctuations by
inflation, we have identified a new variable $\widetilde n_s$, a
function of the basic parameters that can be fixed to great precision
by CMB anisotropy observations.  Distinct models with nearly the same
value of $\widetilde n_s$ cannot be discriminated by CMB data alone.
However, combined with other cosmological observations, determining
$\widetilde n_s$ is a powerful tool for testing models and measuring
fundamental parameters.

We imagine parameterizing the space of cosmological models  by
$$
({\cal C}_2^{(S,T,Is,\ldots)},  \;
   n_{s,t,is,\ldots},\; {\rm h},\; \Omega_B,  \; \Omega_{\Lambda},
\Omega_{CDM},
\;  \Omega_{HDM}, \ldots) \ ,
$$
where $H_0 =
100 \, {\rm h} \, {\rm km \, sec^{-1} Mpc^{-1}}$ is the Hubble parameter,
 and $\Omega_{B,\Lambda, CDM,
HDM,\ldots}$ are the energy densities associated with baryons,
cosmological constant ($\Lambda$), cold and hot dark matter, {\it
etc.}, divided by the critical density.
We use the CMB
quadrupole moments ${\cal C}_2^{(S,T,Is,\ldots)}$ to parameterize the overall
amplitudes of
energy density (scalar metric), gravitational wave (tensor metric),
 isocurvature scalar and other primordial fluctuations predicted by the model.
We parameterize the shape of the initial ({\it e.g.}, post-inflation)
fluctuation spectra in wavenumber $k$
by power law indices $n_{s,t,is,\ldots}$,
defined at time $t_i$ by $k^3
\langle|\widetilde{(\delta \rho/\rho)}(k, t_i)|^2\rangle \propto k^{n_{S}+3}$
and $k^3 \langle|\widetilde{h}_{+, \times}(k, t_i)|^2\rangle \propto
k^{n_{T}}$, where $\delta \rho/\rho$ and $h_{+, \times}$ are the
amplitudes of the energy density and gravitational wave metric
fluctuations (for two polarizations), respectively.

We shall restrict ourselves to subdomains of this large space
consistent with inflation models of fluctuation generation.  Inflation
produces a flat universe, $\Omega_{total} \approx 1$.  We also take
$\Omega_{HDM} =0$, but note that, for angular scales $\gtrsim
10^\prime$, the anisotropy for mixed dark matter models with
$\Omega_{CDM} + \Omega_{HDM} \approx 1$ is quite similar to the
anisotropy if all of the dark matter is cold. Given $\Omega_B$, we
impose the nucleosynthesis estimate [\bbn], $\Omega_B {\rm h}^2 =
0.0125$, to determine ${\rm h}$; we also satisfy globular cluster and
other age bounds [\kolb], and gravitational lens limits [\eturner]: we
range from ${\rm h}\lesssim 0.65$ for $\Omega_{\Lambda}=0$ to ${\rm h}
\lesssim 0.88$ for $\Omega_{\Lambda} \lesssim 0.6$.

Inflation produces adiabatic scalar [\G4] and tensor [\AWSta] Gaussian
fluctuations.  The COBE quadrupole fixes ${\cal C}_2^{(T)}+{\cal
C}_2^{(S)}$, but the tensor-to-scalar quadrupole ratio $r\equiv {\cal
C}_2^{(T)}/{\cal C}_2^{(S)}$ is undetermined (e.g., see Fig.1 in
\Dav).
The indices $n_s$ and $n_t$ are determined by power-law best-fits to
 the theoretical prediction over the scales probed by the CMB.  For
 generic models of inflation, including new, chaotic,
and extended models, inflation gives [\Dav, \Oth, \crit]
\begin{equation}
    n_t \approx n_s-1 \; \; {\rm and}
\; \; r \equiv  {\cal C}_2^{(T)}/{\cal C}_2^{(S)} \approx 7 (1-n_s) \
. \label{eq:ntns}
\end{equation}
Measuring $r$ and $n_s$ to determine whether they respect  Eq.~(\ref{eq:ntns})
is a critical test for inflation.
Exceptions to Eq.~(\ref{eq:ntns})
require
$additional$ fine-tuning of parameters or initial
conditions, beyond that which is strictly necessary
for inflation.  Examples include cosine potentials  (`natural inflation')
 or potentials in any inflationary model  in which an extremum or discontinuity
 is encountered near the end of inflation.
Although we assume Eq.~(\ref{eq:ntns}) for most of our discussion and in our
figures, our calculations are easily extended to these exceptional
inflationary models.  In particular, we have expressed our key
result, Eq.~(\ref{eq:nstilde}), so that it shows the explicit dependence
 on $r$ and $n_s$ separately. Using this expression, the anisotropy
predictions for the exceptional models which violate Eq.~(\ref{eq:ntns})
can be extrapolated from the
results shown in the figures.
With our set of assumptions including Eq.~(\ref{eq:ntns}),
we  have reduced the parameter-space to
three-dimensions, $(r | n_s,{\rm h}, \Omega_{\Lambda})$ (where
$\Omega_B =0.0125 h^{-2}$ and $\Omega_{CDM} = 1-
\Omega_B-\Omega_{\Lambda}$).  We explicitly display both $r$ and $n_s$
but with a ``$|$" as a reminder that $r$ is determined by Eq.~(\ref{eq:ntns})
given $n_s$; we have also assumed $n_t=n_s-1$.

Our results are based on numerical integration of the general
relativistic Boltzmann, Einstein, and hydrodynamic equations for both
scalar [\BE] and tensor metric fluctuations using methods reported
elsewhere [\crit].  Included in the dynamical evolution are all the
relevant components: baryons, photons, dark matter, and massless
neutrinos.  The temperature anisotropy, $\Delta T/T\, (\theta , \phi)=
\sum_{\ell m} a_{\ell m} Y_{\ell m} (\theta,\phi)$, is computed in
terms of scalar and tensor multipole components, $a_{\ell m}^{(S)}$
and $a_{\ell m}^{(T)}$, respectively.  For inflation, each multipole
for the two modes is predicted to be statistically independent and
Gaussian-distributed, fully specified by angular power spectra,
${\cal C}_\ell^{(S)}=\ell(\ell+1) \left\langle \vert a_{\ell m}^{(S)}\vert^2
\right\rangle/(2\pi)$ and
${\cal C}_\ell^{(T)}=\ell(\ell+1) \left\langle \vert a_{\ell m}^{(T)}\vert^2
\right \rangle/(2\pi)$.  \marginpar{Fig. 1}



 Figures 1-4 show sample spectra,
normalized to match the COBE detection. The points on the curves are
weighted averages $\avrg{{\cal C}_\ell}_{W,th}$
of the ${\cal C}_\ell$'s, {\it wrt} weight functions
$W_\ell$:
\begin{equation}
\avrg{{\cal C}_\ell}_{W,th} \equiv  {\cal I}[{\cal C}_\ell W_\ell] /
{\cal I}[ W_\ell] \ , \ {\rm where} \
{\cal I}[f_\ell ] \equiv  \sum_\ell {(\ell+\half)\over \ell
(\ell +1) }\,  f_\ell \  \label{eq:bpow}
\end{equation}
defines the ``logarithmic integral'' ${\cal I}[f_\ell]$ of a function
$f_\ell$. We choose the $W_\ell$ to be filter functions for a set of
existing anisotropy experiments spanning the range $~10^{\circ}$ to
$2'$, some of which report detections in these Proceedings: {\it dmr}
(\dmr), {\it firs} (\mit), {\it ten} (\ten), {\it sp91} (\gaier), {\it
bp} (\sask), {\it pyth} (\python), {\it msam2}, {\it msam3} (\msam),
{\it max} (\max), {\it wd2} (\pwd), {\it ov7} ({\ovro} - {\it ov22} is
a new OVRO experiment).

$\avrg{{\cal C}_\ell}_{W,th}$ characterizes
the broad-band power that the experiment is sensitive to. It is simply
a renormalization by the factor ${\cal I}[W_\ell]$ (typically $\sim
1$) of the {\it rms} fluctuations $(\Delta T/T)_{rms}^2$. The
band-powers are placed at $\avrg{\ell }_W$, and the horizontal bars
(when present) delineate the range that $W_\ell$ covers.  Errors in
the estimation of the band-power $\avrg{{\cal C}_\ell}_{W}$
arise from experimental noise and
the theoretical cosmic variance.  The errors bars shown represent the
limiting resolution achievable with CMB experiments, if there were
full-sky coverage and errors from cosmic variance alone,
\begin{equation}
\avrg{{\cal
C}_\ell}_{W} = \avrg{{\cal C}_\ell}_{W,th} \pm
{ \big\{ {\cal I} [({\cal C}_\ell {W}_\ell)^2/(\ell (\ell +1))]
\big\}^\half \over {\cal I} [ {W}_\ell] } \ .
\end{equation}
The fractional error $\sim \avrg{\ell}^{-1}$ is so tiny for
intermediate and small angle experiments that it would appear that
even extremely subtle differences in the spectra could be determined,
in spite of the $\sim 10\%$ unavoidable error from COBE-type
experiments. Thus in the figures the error bars are much smaller than
the size of the points for $\ell \gta 50$ and for $\ell
\gta 200 $ they basically merge.
For more realistic error
bars, consider a detection obtained from measurements
$(\overline{\Delta T/T})_i \pm \sigma_D$ (where $\sigma_D$ represents
detector noise) at $i=1,\ldots ,N_D$ experimental patches sufficiently
isolated from each other to be largely uncorrelated.  For large $N_D$,
the likelihood function falls by $e^{-\half }$ from a maximum at
$\avrg{{\cal C}_\ell }_{W,maxL}$ when
\begin{eqnarray}
\avrg{{\cal C}_\ell }_{W} =
\avrg{{\cal C}_\ell }_{W,maxL}
  \pm \nu \sqrt{2/N_D}\,  \,
\Big[\avrg{{\cal C}_\ell }_{W,maxL} +
    \sigma_D^2/ {\cal I}[\overline{W}_\ell ]\Big] \, . \label{eq:bpowvar}
\end{eqnarray}
An experimental noise $\sigma_D$ below $10^{-5}$ is standard now, and
a few times $10^{-6}$ is soon achievable; hence, if systematic errors
and unwanted signals can be eliminated, the 1-sigma relative
uncertainty in $\avrg{{\cal C}_\ell }_{W}$ will be from
cosmic-variance alone, $\sqrt{2/N_D}$, falling below $20\%$ for $N_D >
50$.  The observed likelihood maximum is itself centered around
$\avrg{{\cal C}_\ell}_{W,th}$ with a relative error of $1/\sqrt{N_D}$,
but still lies within the error bar, which in fact includes this
effect. The optimal variances shown in the figures roughly correspond
to filling the sky with patches separated by $2\theta_{fwhm}$.

Fig.~ 1 shows how the small-angular signal is increasingly suppressed
as $r$ increases and $n_s$ decreases [\Dav, \crit]. For large maps,
cosmic variance is significant for large-angle experiments [\silk] but
shrinks to insignificant levels at smaller scales.  It appears that $r
| n_s$ could be resolved if $\Lambda$, ${\rm h}$ and ionization
history were known.


Fig. 2 shows the effects of varying $\Omega_{\Lambda}$ or $H_0$
compared to our baseline (solid line) spectrum $(r=0 | n_s=1,{\rm
h}=0.5,\Omega_{\Lambda}=0)$.  Increasing $\Omega_{\Lambda}$ (or
decreasing ${\rm h}$ enhances small-angular scale anisotropy by
reducing the red shift $z_{\rm eq}$ at which radiation-matter equality
occurs.  \marginpar{Fig. 2}
Increasing $\Omega_{\Lambda}$ also changes
slightly the spectral slope for $\ell
\lesssim 10$ due to $\Lambda$-suppression of the growth of scalar
fluctuations [\kofe].  The band-powers show that either $r | n_s$,
$\Omega_{\Lambda}$, or ${\rm h}$ can be resolved if the other two
parameters are known.


A degree of ``cosmic confusion" arises, though, if $r | n_s$,
$\Omega_{\Lambda}$ and ${\rm h}$ vary
simultaneously. \marginpar{Fig. 3}
 Fig. 3 shows spectra for models
lying in a two-dimensional surface of $(r|n_s,{\rm
h},\Omega_{\Lambda})$ which produce nearly identical spectra.  In one
case, $r | n_s$ is fixed, and increasing $\Omega_{\Lambda}$ is nearly
compensated by increasing ${\rm h}$.  In the second case, ${\rm h}$ is
fixed, but increasing $\Omega_{\Lambda}$ is nearly compensated by
decreasing $n_s$ [\kofn]

Further cosmic confusion arises if we consider ionization history.
Let $z_R$ be the red shift at which we suppose sudden, total
reionization of the intergalactic medium.  Fig.~4 compares spectra
with standard recombination (SR), no recombination (NR) and late
reionization (LR) at $z_R=50$, where ${\rm h}=0.5$ and
$\Omega_{\Lambda}=0$.  NR represents the behavior if reionization
occurs early ($z_R>>200$).  The spectrum is substantially suppressed
for $\ell \gtrsim 200$ compared to any SR models.  Experiments at
$\lesssim 0.5^{\circ}$ scale can clearly identify NR or early
reionization ($z_R \gtrsim 150$ gives qualitatively similar results to
NR).  Reionization for $20 \lesssim z_R \lesssim 150$ results in
modest suppression at $\ell \approx 200$, which can be confused with a
decrease in $n_s$ (see figure).  Inflation-inspired models, e.g., cold
dark matter models, are likely to have negligibly small $z_R$ [\BE],
the large $z_R$ examples shown here suggest the small angular-scale
suppression characteristic of models which require large $z_R$, such
as cosmic string and texture models. \marginpar{Fig. 4}

The results can be epitomized by some simple rules-of-thumb: Over the
$30'-2^{\circ}$ range, $\avrg{ {\cal C}_\ell }_{W}$ is roughly
proportional to the maximum of ${\cal C}_\ell$ (the first Doppler
peak).  Since the maximum (corresponding to $\sim .5^{\circ}$ scales)
is normalized to COBE's DMR band-power, $\avrg{ {\cal C}_\ell
}_{dmr}$ (with $W_\ell$ the dmr beam, corresponding to
$\sim 8^{\circ}$ scales), its value is exponentially sensitive
to $n_s$.  Since scalar fluctuations account for the maximum, the
maximum decreases as $r$ increases.  The maximum is also sensitive to
the red shift at matter-radiation equality (or, equivalently,
$(1-\Omega_{\Lambda})h^2$), and to the optical depth at last
scattering for late-reionization models, $\sim z_R^{3/2}$.  These
observations are the basis of an empirical formula (accurate to
$\lesssim 15$\%)
\begin{equation}
\frac{{\cal C}_\ell }{\avrg{ {\cal C}_\ell }_{dmr}}
\Big\vert_{max}
\approx A \; e^{B \; \tilde{n}_s} \ , \label{eq:maxlaw}
\end{equation}
with $A=0.15$, $B=3.56$, and
 \begin{equation}
\begin{array}{rcl}
\tilde{n}_s &  \approx &  n_s -0.28 \, {\rm log}(1+ 0.8 r) \\ & &
-0.52 [(1-\Omega_{\Lambda}){\rm h}^2]^{\frac{1}{2}}\,
    - 0.00036 \, z_R^{3/2}+.26 \ ,  \label{eq:nstilde}
\end{array}
\end{equation}
where $r$ and $n_s$ are related by Eq.~(\ref{eq:ntns})
for generic inflation models,
 and $z_R \lesssim 150$ is needed to have a local maximum.
($\tilde{n}_s$ has been defined such that $\tilde{n}_s=n_s$ for $r=0,\;
h=0.5, \; \Omega_{\Lambda}=0,$ and $z_R=0$.) [In a
forthcoming paper, we show how increasing $\Omega_B h^2$ increases the
Doppler peak and changes the spectral shape.]
Hence, the predicted anisotropy for any experiment
 in the range $10'$ and larger  is not
separately dependent on $n_s$, $r$, $\Omega_\Lambda$, etc.; rather, it
is function of the combination $\tilde{n}_s$.
Eq.~(\ref{eq:nstilde}) shows explicitly the separate dependence on $n_s$
and $r$, and so can be applied to exceptional
inflationary models which violate
Eq.~(\ref{eq:ntns}).


Figure 5 shows how one might use this result, in conjunction with other
astrophysical observations, to determine cosmic parameters.
\marginpar{Fig. 5}
Eq.~(\ref{eq:bpowvar})
implies that the CMB anisotropy measurements are exponentially
sensitive to $\tilde{n}_s$.  Hence, we envisage that $\tilde{n}_s$
will be accurately determined in the foreseeable future.
We suppose, for the purposes of illustration, that experiments
indicate a value $\tilde{n}_s=0.85$.  Then, Eq.~(\ref{eq:nstilde})
implies that
the values of the cosmological parameters are constrained to the
surface illustrated in Fig. 5.
(For simplicity, we have assumed $z_R \lta 20$, as is anticipated
for standard cold dark matter models.)  Cosmological models corresponding
to any point on this surface yield indistinguishable CMB anisotropy
power spectra.  To determine which point on the surface corresponds
to our universe requires other astrophysical measurements.
For example, limits on the age of the universe from globular clusters,
on ${\rm h}$ from Tully-Fisher techniques, on $n_s$ from galaxy and
cluster counts, and on $\Lambda$ from gravitational lenses all reduce
the range of viable parameter space. It is by this combination of
measurements that the CMB power spectrum can develop into a high
precision test of cosmological models.

In the discussion thus far, we have focused on what can be learned from
the CMB anisotropy measurements based on the power spectrum only.
The power spectrum represents only the two-point temperature
correlation function.  From a CMB anisotropy map, one can  hope to
measure three- and higher-point correlation functions, {\it e.g.},
to test for non-gaussianity of the primordial spectrum. Another
conceivable test is the CMB polarization.  Our calculations, though,
suggest that the polarization is unlikely to be detected  or to
provide particularly useful tests of  cosmological parameters [\polar].
For example,  it had been hoped that large-scale (small $\ell$)
polarization measurements would be  useful for discriminating scalar and
tensor contributions to the CMB anisotropy [\hope], thereby
measuring $r$.  In the upper panel
of Fig. 6, we show the percentage polarization (in $\Delta T/T$) for
scalar and tensor modes for a model with $r=1$ and $n_s=.85$, an
example where there are equal tensor and scalar contributions to the
quadrupole moment.  The figure shows that, indeed, there is a dramatically
different polarization expected for scalar versus tensor modes for
small $\ell$.  However, the magnitude of the polarization is less than
0.1\%, probably too small to be detected in the foreseeable future.
On scales less than one degree ($\ell>100$), the total polarization rises and
approaches 10\%,  a more plausible target for detection. However, the
tensor contribution on these angular scales is negligible, so detection
does not permit us to distinguish tensor and scalar modes.  In fact,
the predictions are relatively insensitive to  the cosmological
model, a notable exception being the reionization history.
\marginpar{Fig. 6}
The lower panel of Figure 6 illustrates the prediction for a model
with no recombination.  The overall level of polarization is increased.
The tensor contribution is suppressed relative to scalar, so polarization
remains a poor method of measuring $r$.   However, the polarization
at angular scales of a few degrees ($\ell \approx 50$) rises to nearly
5\%, perhaps sufficient for detection.  An observation of polarization at
these  angular scales would be consistent with a non-standard reionization
history.


We thank  R. Crittenden and G. Efstathiou, who
collaborated in the research from which this summary is drawn.
This research was supported by the DOE at Penn (DOE-EY-76-C-02-3071), NSERC
at Toronto, and the Canadian Institute for Advanced
Research.

\vspace{.2in}

\noindent
{\bf REFERENCES}
\vspace{.1in}

\begin{verse}
Abbott, L. F.,  and Wise, M., 1984,  {\it Nucl.\ Phys.\ B}{\bf 244}, 133.
\re
Adams, F. C., Bond, J. R., Freese, K., Frieman, J. A. and Olinto, A. V.,
1993, {\it  Phys.\ Rev.\ D}{\bf 47}, 426.
\re
Bardeen, J., Steinhardt, P. J., and Turner, M.S., 1983, {\it Phys.
\ Rev. \ Lett.\ D}{\bf 28}, 679.
\re
Bond, J.R.,  Crittenden, J.R.,
Davis, R.L., Efstathiou, G. and Steinhardt, P.J.,
1994,  {\it Phys. Rev. Lett.} {\bf 72}, 13.
\re
Bond, J. R. and Efstathiou, G., 1984, {\it Astrophys.\ J.}{\bf 285} L45;
1987 {\it Mon.\ Not. R.\ astr.\ Soc.} {\bf226}, 655.
\re
Bond, J. R., Efstathiou, G., Lubin, P. M., and Meinhold, P., 1991,
{\it. Phys.\ Rev.\ Lett.} {\bf 66}, 2179.
\re
Cheng, E.S., Meyers, S., and and Page, L.,  1993,   {\it Astrophys. J. Lett.}
{\bf 410} L57.
\re
Crittenden, R., Bond., J.R., Davis., R.L., Efstathiou, G. and Steinhardt, P.J.,
 1993a, {\it Phys.\ Rev.\ Lett.}{\bf 71}, 324.
\re
 Crittenden, R., Davis, R., and Steinhardt, P., 1993b
{\it Astrophys. J.}, L13.
\re
Davis, R. L., Hodges, H. M., Smoot, G. F., Steinhardt, P.J., and Turner, M.S.,
1992, {\it Phys.\ Rev.\ Lett.} {\bf 69}, 1856.
\re
Dragovan, M. et al., 1993, Berkeley AAS  1993 Meeting.
\re
Gaier, T., Schuster, J., Gunderson, J., Koch, T., Seiffert, M.,
Meinhold, P., and Lubin, P. 1992, {\it Astrophys.\ J.\ Lett.} {\bf 398} L1.
\re
Ganga, K., {\it et.\ al.}, 1993,  preprint.
\re
Gunderson, J., {\it et \ al.}, 1993,  {\it Astrophys.\ J.\ Lett.} {\bf 413},
L1.
\re
Guth, A.H., 1981, {\it Phys.\ Rev.\ D} {\bf 23}, 347.
\re
Guth, A.H., and Pi, S.-Y., 1982, {\it Phys.\ Rev.\ Lett.} {\bf 49}, 1110.
\re
Hawking, S.W., 1982, {\it Phys.\ Lett.\ B}{\bf 115}, 295.
\re
Kofman, L., Bahcall, Gnedin, N., and Bahcall, N.,  1993,
 {\it Astrophys.J.} {\bf  413}, 1.
\re
Kofman, L. and Starobinskii, A.A., 1986, {\it Sov.\ Astron.\ Lett.} {\bf 11},
			271.
\re
Liddle, A. and Lyth, D., 1992, {\it Phys.\ Lett.\ B}{\bf 291}, 391.
\re
Lidsey, J.E. and Coles, P., 1992, {\it Mon.\ Not.\ Roy.\ astr.\ Soc.} {\bf 258}
57P.
\re
Lucchin, R., Matarrese, S., and Mollerach, S., 1992, {\it Astrophys.\ J.\
Lett.} {\bf 401}, 49.
\re
Meinhold, P. and Lubin, P., 1991, {\it Astrophys.\
J.\ Lett.} {\bf 370}, 11.
\re
Meinhold, P.,  {\it. et.\ al.}, 1993,   {\it Astrophys.\ J.\ Lett.} {\bf 409},
L1.
\re
Polnarev, A.G., 1985, {\it Sov.\ Astron.} {\bf 29}, 607.
\re
Readhead, A.C.S., 1989, et al., OVRO (Owens Valley), {\it Astrophys.\ J.}\
{\bf 346}, 556.
\re
Rubakov, V. A., Sazhin, M. V., Veryaskin, A. V., {\it Phys. \  Lett.
\ B}{\bf 115},
189.
\re
Salopek, D., 1992, {\it Phys.\ Rev.\ Lett.} {\bf 69}, 3602.
\re
Schuster, J., et al., 1993, {\it Astrophys.\ J.\ Lett.} {\bf 412}, L47.
\re
Smoot, G.F., et al., 1992 {\it Astrophys.\ J.\ Lett.} {\bf 396}, L45.
\re
Starobinskii, A.A., 1982, {\it. Phys.\ Lett.\ B}{\bf 117}, 175.
\re
Starobinskii, A.A., 1985, {\it Sov.\ Astron.\ Lett.} {\bf 11}, 133.
\re
Tucker, G.S.,   Griffin, G.S.,  Nguyen, H.  and  Peterson, J.B. 1993, .
{\it Astrophys.\ J.\ Lett.} {\bf 419}, L45.
\re
Walker, T.P., Steigman, G., Schramm, D.N., Olive, K.A., and
Kang, H.S., 1991, {\it Astrophys.\ J.} {\bf 376}, 5l.
\re
Watson R., Guteirrez de la Cruz, C.M., Davies, R., Lasenby, A.N.,
Rebolo, R., Beckman, J.E., and Hancock, S., 1992, {\it Nature} {\bf 357}
660.
\re
White, M., Silk, J., and Krauss, L.,  1993, {\it Astrophys.\ J.\
Lett.}
{\bf 418}, 535.
\re
Wollack, E.J.,  Jarosik, N.C.,  Netterfield, C.B.,  Page, L.A., and
  Wilkinson, D.,  1993, Princeton preprint.
\end{verse}

\newpage

\begin{center}
{\bf FIGURE CAPTIONS}
\end{center}
\vspace{.2in}

\noindent
{\bf Figure 1.}  Power spectra as a function of multipole moment
$\ell$ for ($r$=$0 | n_s$=$1$), ($r$=$0.7 | n_s$=$0.9$) and ($r$=$1.4
| n_s$=$0.8$) where ${\rm h}=0.5 \; {\rm and} \; \Omega_{\Lambda}=0$
for all models.  The spectra in all figures are normalized so that the
COBE band-power is $10^{-10}$. It is observed to be
$(1.0\pm 0.2) \times 10^{-10}$. The (elongated) vertical bars on
the band-powers $\avrg{{\cal C}_\ell}_{W}$ are 1-sigma
cosmic variance error bars  assuming full-sky
coverage for the 11 experiments shown (see also Eq.~\ref{eq:bpowvar}).
The band-powers are placed at the average $\avrg{\ell}_{W}$ of
$\ell$ over the filter. The horizontal error bars on the $n_s$=$1$ case
give the ranges where the filters fall by
half an e-folding from their maxima. \\ \\

\noindent
{\bf Figure 2.}
Power spectra as a function of $\ell$ for
scale-invariant models, with $r=0 | n_s=1$.  The middle curve shows
${\rm h}=0.5$ and $\Omega_{\Lambda} =0$. In the upper
curve, $\Omega_{\Lambda}$ is increased to 0.4 while keeping ${\rm
h}=0.5$.  In the lower  curve, $\Omega_{\Lambda}=0$ but ${\rm
h}$ is increased from 0.5 to 0.65 (hence $\Omega_B$ drops from $0.5$ to
$0.3$).  The spectra are insensitive to changes in ${\rm h}$ for fixed
$\Omega_B$. Increasing $\Omega_{\Lambda}$ or $\Omega_B$ increases the
power at $\ell \sim 200$.  \\ \\

\noindent
{\bf Figure 3.}
Examples of different cosmologies with nearly
identical spectra of multipole moments and values of the band-powers.
The solid curve is $(r=0 | n_s=1,{\rm h}=0.5, \Omega_{\Lambda}=0)$.  The
other two curves explore degeneracies in the $(r=0 | n_s=1,{\rm
h},\Omega_{\Lambda})$ and $(r | n_s,{\rm h}=0.5,\Omega_{\Lambda})$
planes.  In the dashed curve, increasing $\Omega_{\Lambda}$ is almost
exactly compensated by increasing ${\rm h}$.  In the dot-dashed curve,
the effect of changing to $r=0.42 | n_s=0.94$ is nearly compensated by
increasing $\Omega_{\Lambda}$ to 0.6. \\ \\

\noindent
{\bf Figure 4.}
Power spectra for models with standard recombination (SR), no
recombination (NR), and `late' reionization (LR) at $z=50$.  In all
models, ${\rm h}=0.5$ and $\Omega_{\Lambda}=0$.  NR or
reionization at $z \ge 150$ results in substantial suppression at
$\ell \ge 100$.  Models with reionization at $20 \le z \le 150$
give moderate suppression that can mimic decreasing $n_s$ or
increasing ${\rm h}$; {\it e.g.}, compare the $n_s=0.95$ spectrum with SR
(thin, dot-dashed) to the $n_s=1$ spectrum with reionization at $z=50$
(thick, dot-dashed). \\   \\

\noindent
{\bf Figure 5.}
The surface in the parameter-space $(n_s, \; h,\; \Omega_{\Lambda})$
corresponding
to $\tilde{n}_s=0.85$, as determined by Eq.~(\ref{eq:nstilde}).
The grey-scale varies with the height of the surface, or,
equivalently,
the value of the spectral
tilt $n_s$, the darkest strip corresponds to $n_s \approx 0.85$.
Note that  $\tilde{n}_s=0.85$ is also consistent with values of $n_s$
quite different from 0.85.
Choices of cosmological parameters corresponding
each point along the surface yield  virtually indistinguishable CMB
anisotropy.  \\ \\

\noindent
{\bf Figure 6.}
The percentage polarization in $\Delta T/T$  versus multipole
moment $\ell$ predicted for
  an inflationary model with $n_s=0.85$, $h=0.5$, cold
  dark matter, and standard
   recombination.  (For this value of
$n_s$, inflation predicts equal scalar and tensor
    contributions to the unpolarized quadrupole.)
The upper panel represents the prediction  for standard recombination and
the lower panel is for a model with no recombination.

\end{document}